\documentclass[nofootinbib,prd,aps,superscriptaddress,preprintnumbers,twocolumn,showpacs]{revtex4-1}
\usepackage[dvipdfmx]{graphicx}
\usepackage{amsmath,amssymb,amsthm,bm,bigdelim,multirow}
\usepackage{color}

\newcommand{\kslash}{k\kern-1ex /}
\newcommand{\pslash}{p\kern-1ex /}
\newcommand{\qslash}{q\kern-1ex /}
\newcommand{\lslash}{l\kern-1ex /}
\newcommand{\sslash}{s\kern-1ex /}
\newcommand{\Dslash}{D\kern-1.2ex /}

\newcommand{\beqa}{\begin{eqnarray}}
\newcommand{\eeqa}{\end{eqnarray}}

\newcommand{\bd}{\begin{description}}
\newcommand{\ed}{\end{description}}

\newcommand{\ben}{\begin{eqnarray}}
\newcommand{\een}{\end{eqnarray}}

\def\lsim{\raise0.3ex\hbox{$<$\kern-0.75em\raise-1.1ex\hbox{$\sim$}}}
\def\gsim{\raise0.3ex\hbox{$>$\kern-0.75em\raise-1.1ex\hbox{$\sim$}}}
\def\simgt{\rlap{\lower 3.5 pt\hbox{$\mathchar \sim$}}\raise 2.0pt \hbox {$>$}}
\def\simlt{\rlap{\lower 3.5 pt\hbox{$\mathchar \sim$}}\raise 2.0pt \hbox {$<$}}

\begin{document}
\title{Calculation of fermionic Green functions\\
with Grassmann higher-order tensor renormalization group}

\author{Yusuke Yoshimura}
\affiliation{Center for Computational Sciences, University of Tsukuba, Tsukuba, Ibaraki 305-8577, Japan}

\author{Yoshinobu Kuramashi}
\affiliation{Center for Computational Sciences, University of Tsukuba, Tsukuba, Ibaraki 305-8577, Japan}
\affiliation{RIKEN Advanced Institute for Computational Science, Kobe, Hyogo 650-0047, Japan}

\author{Yoshifumi Nakamura}
\affiliation{RIKEN Advanced Institute for Computational Science, Kobe, Hyogo 650-0047, Japan}
\affiliation{Graduate School of System Informatics, Department of Computational Sciences, Kobe University, Kobe, Hyogo 657-8501, Japan}

\author{Shinji Takeda}
\affiliation{Institute for Theorical Physics, Kanazawa University, Kanazawa 920-1192, Japan}

\author{Ryo Sakai}
\affiliation{Institute for Theorical Physics, Kanazawa University, Kanazawa 920-1192, Japan}

\begin{abstract}
We develop calculational method for fermionic Green functions in the framework of Grassmann higher-order tensor renormalization group. The validity of the method is tested by applying it to three-dimensional free Wilson fermion system. We compare the numerical results for chiral condensate and two-point correlation functions with the exact ones obtained by analytical methods.
\end{abstract}
\date{\today}

\preprint{UTHEP-708, UTCCS-P-108, KANAZAWA-17-09}

\maketitle

\newpage
\section{Introduction}
The tensor renormalization group (TRG) is a numerical renormalization group method in the tensor network scheme, which was originally introduced by Levin and Nave~\cite{trg}.
This method consists of two steps. We first make a tensor network representation for the target physical system.
After that, we repeat the coarse-graining of the tensor network based on the singular value decomposition (SVD).
The basic concept of this local transformation is similar to the block spin transformation~\cite{block}.
An advantage of the TRG method is that it does not suffer from the sign problem inherent in the Monte Carlo methods.
Subsequently to the appearance of TRG, it was extended for Grassmann valued tensor networks, and the new method is called Grassmann TRG (GTRG)~\cite{gtrg,Schwinger1}.
The TRG and GTRG methods have been applied to the two-dimensional relativistic quantum field theories in elementary particle physics:
lattice $\phi^4$ model~\cite{phi4}, lattice Schwinger model~\cite{Schwinger1,Schwinger2}, and the finite-density lattice Thirring model~\cite{Thirring}.
Meanwhile, the TRG method itself has been improved by disentangling short-range correlations in tensor network renormalization (TNR)~\cite{tnr} and loop-TNR~\cite{loopTNR}.

Although the original idea of TRG was restricted to two-dimensional systems, Xie {\it et al.} invented a new method called higher-order TRG (HOTRG) with the use of higher-order SVD (HOSVD), which is formally applicable to any higher-dimensional system~\cite{hotrg}.
Its effectiveness is also demonstrated in Ref.~\cite{hotrg} by applying it to two- and three-dimensional Ising models.
So far, the HOTRG method has been applied to several two- and three-dimensional systems:
three-dimensional Potts models~\cite{Potts}, two-dimensional XY model~\cite{XY}, two-dimensional O(3) model~\cite{O3}, and two-dimensional CP(1) model~\cite{CP1}.

In order to use the tensor network scheme in lattice QCD, we need an algorithm which can be applied to four-dimensional relativistic fermion systems.
The authors in Ref.~\cite{ghotrg} have formulated the Grassmann HOTRG (GHOTRG) based on the GTRG.
It opened the way to higher-dimensional fermion systems but did not allow us to calculate physical quantities except the free energy and its differentiation.
In this article we develop a method for direct calculation of fermionic Green functions in the framework of GHOTRG and test the validity of the algorithm using the relativistic free Wilson fermion system, for which the exact values of Green functions are analytically calculable.

This paper is organized as follows.
We make a brief review for the GHOTRG algorithm in Sec.~\ref{sec:GHOTRG}. 
In Sec.~\ref{sec:Gfunction} we explain calculational method of fermionic Green functions with the GHOTRG method. Numerical results for Green functions in the relativistic free Wilson fermion system are presented in Sec.~\ref{sec:results}. 
Section~\ref{sec:summary} is devoted to summary and outlook.

\section{Grassmann higher-order tensor renormalization}
\label{sec:GHOTRG}

The detailed explanation for the GHOTRG algorithm is already given for the two-dimensional case in
Ref.~\cite{ghotrg}.
Although the extension to three-dimensional cases is rather straightforward, we give a brief review to fix the notations.
We assume $a=1$ for the lattice units in the following.

\subsection{Tensor network representation}
We first present a tensor network representation of the partition function for
the three-dimensional free Wilson fermion with the Wilson parameter $r=1$ which is described in detail in Ref.~\cite{ghotrg}.
The action with the fermion mass $m$ is given by
\begin{gather}
	S=
	\sum_{n,n^\prime} \bar\psi_n D_{n,n^\prime} \psi_{n^\prime}, \\
	D_{n,n^\prime}=
	(m+3) \delta_{n,n^\prime}
	-\frac{1}{2} \sum_{\mu,\pm} \left( 1\pm\gamma_\mu \right) \delta_{n,n^\prime\pm\hat\mu}
\end{gather}
where $n=(n_1,n_2,n_3)$ is the lattice coordinate and the fermion fields $\psi_n,\bar\psi_n$ have two spinor components.
A tensor network representation of the partition function is written by
\begin{gather}
	Z=
	\int \mathcal D\psi \mathcal D\bar\psi\ e^{-S}=
	\sum_{\{x=0,1\}} \int\prod_n \mathcal T_{ l_n }, \\
	l_n	=
	x_{1,n} \otimes x_{2,n} \otimes x_{3,n} \otimes x_{1,n-\hat 1} \otimes x_{2,n-\hat 2} \otimes x_{3,n-\hat 3}
\end{gather}
where the indices are given by
\begin{equation}
	x_{i,n}=\left( x_{i,n,1},x_{i,n,2} \right),\ i=1,2,3,
\end{equation}
representing that $\psi$ has two spinor components.
The tensor $\mathcal T$ is defined as
\begin{equation}
	\mathcal T_{ l_n }= T_{ l_n } G_{ l_n }
	\label{eq:tnr_tensor}
\end{equation}
where the bosonic tensor $T$ have normal number elements whose explicit form is given in Ref.~\cite{ghotrg},
and the Grassmann tensor $G$ is defined as
\begin{align}
	G_{ l_n }&=
	\left( \prod_{i=1}^3 d\bar\eta_{i,n,2}^{x_{i,n,2}} d\eta_{i,n,1}^{x_{i,n,1}} \right)
	\left( \prod_{i=1}^3 d\eta_{i,n,2}^{x_{i,n-\hat i,2}} d\bar\eta_{i,n,1}^{x_{i,n-\hat i,1}} \right)
	\nonumber \\ &\times
	\prod_{i=1}^3
	\left( \bar\eta_{i,n+\hat i,1}\eta_{i,n,1} \right)^{x_{i,n,1}}
	\left( \bar\eta_{i,n,2}\eta_{i,n+\hat i,2} \right)^{x_{i,n,2}},
	\\ &\qquad
	\int d\eta_{i,n,s} \eta_{i,n,s}=
	\int d\bar\eta_{i,n,s} \bar\eta_{i,n,s}= 1.
\end{align}

The original fermion fields $\psi,\bar\psi$ have already been integrated out, while another set of independent Grassmann variables $\eta,\bar\eta$ have been introduced.
Because of the Grassmann nature,
elements of the bosonic tensor take non-trivial values only when the indices satisfy the following condition:
\begin{equation}
	\left[ \sum_{i=1}^3 \sum_{s=1}^2 \left( x_{i,n,s}+x_{i,n-\hat i,s} \right) \right] \bmod 2=0.
	\label{eq:modT}
\end{equation}
Otherwise the element is zero. This allows us the block diagonalization of the bosonic tensors explained in Appendix A, which contributes to save the computational and memory costs.

\subsection{Renormalization of bosonic tensor}
\begin{figure}
	\centering
	\includegraphics[scale=0.4]{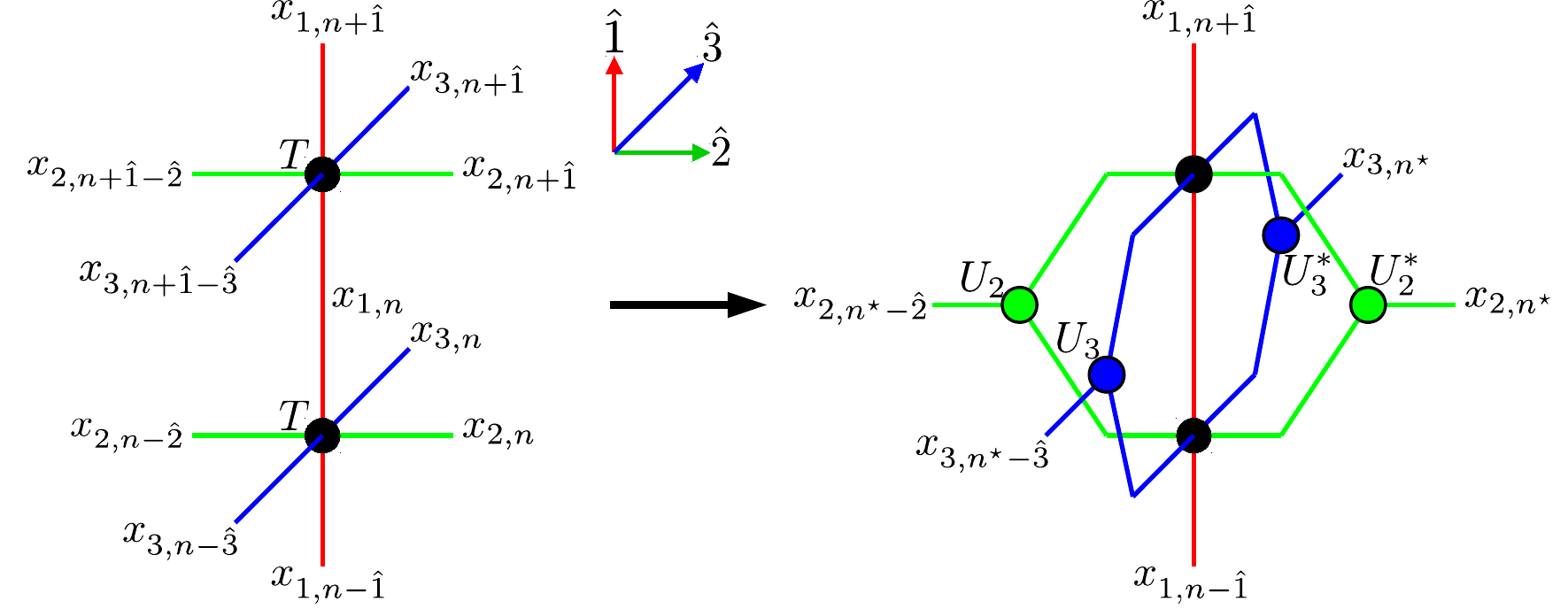}
	\caption{Coarse-graining along $x_1$-direction.}
	\label{fig:hotrg3}
\end{figure}

We explain the renormalization procedure of bosonic tensors, which is mostly based on the original HOTRG~\cite{hotrg}.
Here we focus on the coarse-graining along $\hat 1$-direction depicted in Fig.~\ref{fig:hotrg3}.
The coarse-graining for the other directions can be done in a similar way.

We first make up a contracted tensor $M$ by
\begin{gather}
	M_{ m_n }=
	\sum_{ x_{1,n} } T_{ l_{n+\hat 1} } T_{ l_n } \sigma_{ l_{n+\hat 1} l_n },
	\label{eq:tensorM} \\
	m_n=
	x_{1,n+\hat 1} \otimes x_{2,n}^+ \otimes x_{3,n}^+ \otimes x_{1,n-\hat 1} \otimes x_{2,n}^- \otimes x_{3,n}^-
\end{gather}
where the integrated indices are defined as
\begin{align}
	x_{i,n}^+
	&=x_{i,n}\otimes x_{i,n+\hat 1},\\
	x_{i,n}^-
	&=x_{i,n-\hat i}\otimes x_{i,n+\hat 1-\hat i}
\end{align}
for $i=2,3$.
$\sigma$ is a sign factor originating from anticommutation properties in the renormalization of Grassmann part, which is explained in the next subsection.

\begin{figure}
	\centering
	\includegraphics[scale=0.4]{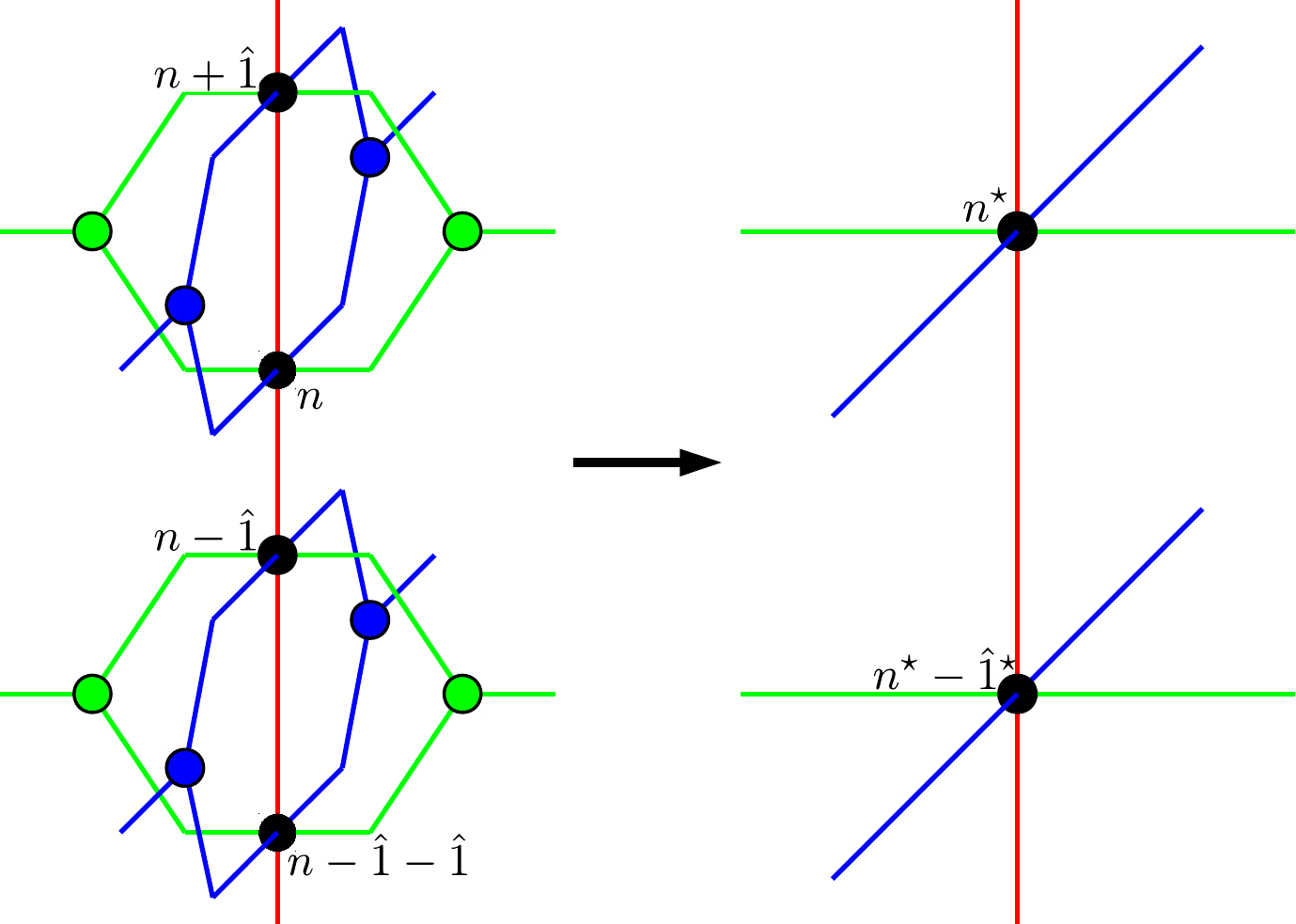}
	\caption{Update of indices from $n$ to $n^\star$.}
	\label{fig:hotrg3star}
\end{figure}

Next we apply the eigenvalue decomposition to $M^+$ and $M^-$ defined as
\begin{align}
	\left( M^+ \right)_{x_{2,n}^+,x_{2,n}^{+\prime}}
	&=\sum_{a,c,d,e,f} M_{ a x_{2,n}^+ cdef } M_{ a x_{2,n}^{+\prime} cdef }^\ast, \\
	\left( M^- \right)_{x_{2,n}^-,x_{2,n}^{-\prime}}
	&=\sum_{a,b,c,d,f} M_{ abcd x_{2,n}^- f } M_{ abcd x_{2,n}^{-\prime} f }^\ast
\end{align}
and obtain a unitary matrix $U^\pm$ and the corresponding eigenvalues $\lambda^\pm$:
\begin{equation}
	\left( M^{\pm} \right)_{ x_{2,n}^\pm,x_{2,n}^{\pm\prime} }=
	\sum_k \left(U^\pm\right)_{ x_{2,n}^\pm k } \lambda^\pm_k \left(U^\pm\right)^\ast_{ x_{2,n}^{\pm\prime} k }.
	\label{eq:evd}
\end{equation}
Then, while assuming that $\lambda^\pm_i$ are obtained in descending order, we define $\epsilon^\pm$ as
\begin{align}
	\epsilon^\pm= \sum_{i>D_{\rm{cut}}} \lambda_i^\pm
	\label{eq:epsilon}
\end{align}
with a given $D_{\rm{cut}}$ that one can choose.
If $\epsilon^+ < \epsilon^-$ then $U_2=U^+$ is adopted as an isometory to be used for the coarse-graining,
and vice versa.
In a similar manner an isometory $U_3$ for $\hat 3$-direction is obtained.
Restricting the new indices $1\leq x^{\rm{b}}_{2,n^\star},x^{\rm{b}}_{3,n^\star} \leq D_{\rm{cut}}$,
the coarse-grained bosonic tensor $T^\star$ is defined as a product of $U_2,U_3$ and $M$:
\begin{align}
	T^{\star,\rm{b}}_{ l^{\rm{b}}_{n^\star} }
	&=\sum_{ x_{2,n}^\pm,x_{3,n}^\pm }
	M_{m_n}
	\prod_{i=2}^3
	(U_i)_{ x_{i,n}^+ x^{\rm{b}}_{i,n^\star} }^\ast (U_i)_{ x_{i,n}^- x^{\rm{b}}_{i,n^\star-\hat i} },
	\\
	l^{\rm{b}}_{n^\star}
	&=x_{ 1,n^\star } \otimes
	x^{\rm{b}}_{ 2,n^\star} \otimes
	x^{\rm{b}}_{ 3,n^\star }
	\nonumber \\ &\qquad\otimes
	x_{ 1,n^\star-\hat 1^\star } \otimes
	x^{\rm{b}}_{ 2,n^\star-\hat 2 } \otimes
	x^{\rm{b}}_{ 3,n^\star-\hat 3 }.
	\label{eq:tensorTstar}
\end{align}
For the lattice coordinate of the index $l$ for the tensor,
here we unify $n,n+\hat 1$ to $n^\star$ and define the new unit vector as $\hat 1^\star=\hat 1+\hat 1$ as shown in Fig.~\ref{fig:hotrg3star}.

\subsection{Renormalization of Grassmann part}
We now explain the renormalization procedure of the Grassmann part focusing on the coarse-graining along $\hat 1$-direction.

We first extract the Grassmann parts:
\begin{widetext}
\begin{multline}
	G_{ l_{n+\hat 1} } G_{ l_n }=
	\left(
		\prod_{i=1}^3
		d\bar\eta_{i,n+\hat 1,2}^{x_{i,n+\hat 1,2}}
		d\eta_{i,n+\hat 1,1}^{x_{i,n+\hat 1,1}}
	\right)
	\left(
		\prod_{i=1}^3
		d\eta_{i,n+\hat 1,2}^{ x_{i,n+\hat 1-\hat i,2} }
		d\bar\eta_{i,n+\hat 1,1}^{ x_{i,n+\hat 1-\hat i,1}}
	\right)
	\left( \prod_{i=1}^3 d\bar\eta_{i,n,2}^{x_{i,n,2}} d\eta_{i,n,1}^{x_{i,n,1}} \right)
	\left( \prod_{i=1}^3 d\eta_{i,n,2}^{x_{i,n-\hat i,2}} d\bar\eta_{i,n,1}^{x_{i,n-\hat i,1}} \right) \\
	\times \prod_{i=1}^3
	\left( \bar\eta_{i,n+\hat 1+\hat i,1} \eta_{i,n+\hat 1,1} \right)^{x_{i,n+\hat 1,1}}
	\left( \bar\eta_{i,n+\hat 1,2} \eta_{i,n+\hat 1+\hat i,2} \right)^{x_{i,n+\hat 1,2}}
	\left( \bar\eta_{i,n+\hat i,1} \eta_{i,n,1} \right)^{x_{i,n,1}}
	\left( \bar\eta_{i,n,2} \eta_{i,n+\hat i,2} \right)^{x_{i,n,2}}.
	\label{eq:masures}	
\end{multline}
\end{widetext}
Next, in order to control sign factors from the anticommutation of Grassmann numbers on other sites,
we introduce a new index and a new set of Grassmann variables satisfying
\begin{align}
	\left( d\eta_{i,n^\star-\hat i}d\bar\eta_{i,n^\star}\bar\eta_{i,n^\star}\eta_{i,n^\star-\hat i} \right)
	^{ x^{\rm{f}}_{i,n^\star-\hat i} }=1,\quad i=2,3
	\label{eq:etaStar}
\end{align}
where new indices are defined as
\begin{align}
	x^{\rm{f}}_{i,n^\star-\hat i}
	=\left[ \sum_{s=1}^2 \left( x_{i,n-\hat i,s}+x_{i,n+\hat 1-\hat i,s} \right) \right] \bmod 2.
	\label{eq:xf}
\end{align}
By inserting Eq.~(\ref{eq:etaStar}) to Eq.~(\ref{eq:masures}) and integrating
$\bar\eta_{ 1,n+\hat 1,1 },\eta_{1,n,1},\bar\eta_{1,n,2}$
and $\eta_{ 1,n+\hat 1,2 }$,
we obtain
\begin{widetext}
\begin{multline}
	(-1)^{
		x_{1,n,1} \left( 1+x_{i,n,2} \right)+
		\left( x_{2,n+\hat 1-\hat 2,1}+x_{2,n+\hat 1-\hat 2,2} \right)
		\left( x_{3,n-\hat 3,1}+x_{3,n-\hat 3,2} \right)
	}
	\left(
		\prod_{i=1}^3
		d\bar\eta_{i,n+\hat 1,2}^{x_{i,n+\hat 1,2}}
		d\eta_{i,n+\hat 1,1}^{x_{i,n+\hat 1,1}}
	\right)
	\left( \prod_{i=2}^3 d\bar\eta_{i,n,2}^{x_{i,n,2}} d\eta_{i,n,1}^{x_{i,n,1}} \right)
	\\ \times
	d\eta_{1,n,2}^{x_{1,n-\hat 1,2}} d\bar\eta_{1,n,1}^{x_{1,n-\hat 1,1}}
	\left[
		\prod_{i=2}^3
		d\eta_{i,n,2}^{x_{i,n-\hat i,2}}
		d\bar\eta_{i,n,1}^{x_{i,n-\hat i,1}}
		d\eta_{i,n+\hat 1,2}^{ x_{i,n+\hat 1-\hat i,2} }
		d\bar\eta_{i,n+\hat 1,1}^{ x_{i,n+\hat 1-\hat i,1}}
		d\eta_{i,n^\star-\hat i}^{ x^{\rm{f}}_{i,n^\star-\hat i} }
		d\bar\eta_{i,n^\star}^{ x^{\rm{f}}_{i,n^\star-\hat i} }
		\left( \bar\eta_{i,n^\star}\eta_{i,n^\star-\hat i} \right)^{ x^{\rm{f}}_{i,n^\star-\hat i} }
	\right]
	\\ \times
	\left[
		\prod_{i=1}^3
		\left( \bar\eta_{i,n+\hat 1+\hat i,1} \eta_{i,n+\hat 1,1} \right)^{x_{i,n+\hat 1,1}}
		\left( \bar\eta_{i,n+\hat 1,2} \eta_{i,n+\hat 1+\hat i,2} \right)^{x_{i,n+\hat 1,2}}
	\right]
	\left[
		\prod_{i=2}^3
		\left( \bar\eta_{i,n+\hat i,1} \eta_{i,n,1} \right)^{x_{i,n,1}}
		\left( \bar\eta_{i,n,2} \eta_{i,n+\hat i,2} \right)^{x_{i,n,2}}
	\right].
	\label{eq:gparts}
\end{multline}
\end{widetext}
Thanks to Eq.~(\ref{eq:xf}),
the combinations of Grassmann measures
\[
	d\eta_{i,n,2}^{x_{i,n-\hat i,2}}
	d\bar\eta_{i,n,1}^{x_{i,n-\hat i,1}}
	d\eta_{i,n+\hat 1,2}^{ x_{i,n+\hat 1-\hat i,2} }
	d\bar\eta_{i,n+\hat 1,1}^{ x_{i,n+\hat 1-\hat i,1}}
	d\eta_{i,n^\star-\hat i}^{ x^{\rm{f}}_{i,n^\star-\hat i} } 
\]
for $i=2,3$ are Grassmann-even and thus we can now freely move it.
By shifting the coordinate of the Grassmann measures and
$\bar\eta_{2,n^\star}\eta_{2,n^\star-\hat 2},\bar\eta_{3,n^\star}\eta_{3,n^\star-\hat 3}$,
one find that Eq.~(\ref{eq:gparts}) turns out to be
\begin{widetext}
\begin{multline}
	(-1)^{
		x_{1,n,1} \left( 1+x_{i,n,2} \right)+
		\left( x_{2,n+\hat 1-\hat 2,1}+x_{2,n+\hat 1-\hat 2,2} \right)
		\left( x_{3,n-\hat 3,1}+x_{3,n-\hat 3,2} \right)+
		x^{\rm{f}}_{2,n^\star} x^{\rm{f}}_{3,n^\star}
	}
	d\bar\eta_{1,n+\hat 1,2}^{x_{1,n+\hat 1,2}}
	d\eta_{1,n+\hat 1,1}^{x_{1,n+\hat 1,1}}
	\\ \times
	\left\{
	\left(
		\prod_{i=2}^3
		d\bar\eta_{i,n+\hat 1,2}^{x_{i,n+\hat 1,2}}
		d\eta_{i,n+\hat 1,1}^{x_{i,n+\hat 1,1}}
	\right)
	\left(
		\prod_{i=2}^3
		d\bar\eta_{i,n,2}^{x_{i,n,2}}
		d\eta_{i,n,1}^{x_{i,n,1}}
	\right)
	\left(
		\prod_{i=2}^3
		d\eta_{i,n+\hat i,2}^{x_{i,n,2}}
		d\bar\eta_{i,n+\hat i,1}^{x_{i,n,1}}
		d\eta_{i,n+\hat 1+\hat i,2}^{ x_{i,n+\hat 1,2} }
		d\bar\eta_{i,n+\hat 1+\hat i,1}^{ x_{i,n+\hat 1,1}}
	\right)
	\right.
	\\ \times
	\left.
	\left[
		\prod_{i=2}^3
		\left( \bar\eta_{i,n+\hat 1+\hat i,1} \eta_{i,n+\hat 1,1} \right)^{x_{i,n+\hat 1,1}}
		\left( \bar\eta_{i,n+\hat 1,2} \eta_{i,n+\hat 1+\hat i,2} \right)^{x_{i,n+\hat 1,2}}
		\left( \bar\eta_{i,n+\hat i,1} \eta_{i,n,1} \right)^{x_{i,n,1}}
		\left( \bar\eta_{i,n,2} \eta_{i,n+\hat i,2} \right)^{x_{i,n,2}}
	\right]
	\right\}
	\\ \times
	\left( \prod_{i=2,3} d\eta_{i,n^\star}^{ x^{\rm{f}}_{i,n^\star} } \right)
	d\eta_{1,n,2}^{x_{1,n-\hat 1,2}} d\bar\eta_{1,n,1}^{x_{1,n-\hat 1,1}}
	\left( \prod_{i=2,3} d\bar\eta_{i,n^\star}^{ x^{\rm{f}}_{i,n^\star-\hat i} } \right)
	\\ \times
	\left( \bar\eta_{1,n+\hat 1+\hat 1,1} \eta_{1,n+\hat 1,1} \right)^{x_{1,n+\hat 1,1}}
	\left( \bar\eta_{1,n+\hat 1,2} \eta_{1,n+\hat 1+\hat 1,2} \right)^{x_{1,n+\hat 1,2}}
	\prod_{i=2,3} \left( \bar\eta_{i,n^\star+\hat i}\eta_{i,n^\star} \right)^{ x^{\rm{f}}_{i,n^\star} },
\end{multline}
\end{widetext}
where the Grassmann numbers in the braces can be integrated out.
After the integration, the coarse-grained Grassmann part and the sign factor of Eq.~(\ref{eq:tensorM}) results in
\begin{widetext}
\begin{align}
	&\sigma_{ l_{n+\hat 1}l_n }G_{ l^{\rm{f}}_{n^\star} },
	\quad l^{\rm{f}}_{n^\star}=
	x_{1,n^\star} \otimes x^{\rm{f}}_{2,n^\star} \otimes x^{\rm{f}}_{3,n^\star}
	\otimes x_{1,n^\star-\hat 1^\star}
	\otimes x^{\rm{f}}_{2,n^\star-\hat 2}
	\otimes x^{\rm{f}}_{3,n^\star-\hat 3},
	\\ &
	G^\star_{ l^{\rm{f}}_{n^\star} }=
	d\bar\eta_{ 1,n^\star,2 }^{ x_{1,n^\star,2} }
	d\eta_{ 1,n^\star,1 }^{ x_{1,n^\star,1} }
	\left( \prod_{i=2}^3 d\eta_{i,n^\star}^{x^{\rm{f}}_{i,n^\star}} \right)
	d\eta_{1,n^\star,2}^{x_{1,n^\star-\hat 1^\star,2}}
	d\bar\eta_{1,n^\star,1}^{x_{1,n^\star-\hat 1^\star,1}}
	\left( \prod_{i=2}^3 d\bar\eta_{i,n^\star}^{x^{\rm{f}}_{i,n^\star-\hat i}} \right)
	\nonumber \\ &\qquad\qquad\times
	\left( \bar\eta_{1,n^\star+\hat 1^\star,1} \eta_{1,n^\star,1} \right)^{ x_{1,n^\star,1} }
	\left( \bar\eta_{1,n^\star,2} \eta_{1,n^\star+\hat 1^\star,2} \right)^{ x_{2,n^\star,2} }
	\prod_{i=2}^3
	\left( \bar\eta_{i,n^\star+\hat i} \eta_{i,n^\star} \right)^{ x^{\rm{f}}_{i,n^\star} },
	\\ &
	\sigma_{ l_{n+\hat 1}l_n }
	=(-1)
	^{
		x_{1,n,1} \left( 1+x_{1,n,2} \right)
		+\left( x_{2,n+\hat 1-\hat 2,1}+x_{2,n+\hat 1-\hat 2,2} \right) \left( x_{3,n-\hat 3,1}+x_{3,n-\hat 3,2} \right)
		+\left( x_{2,n,1}+x_{2,n,2} \right) \left( x_{3,n+\hat 1,1}+x_{3,n+\hat 1,2} \right)
	}
	\nonumber\\&\qquad\qquad
	\times(-1)^{
		x_{2,n,2}\left( x_{2,n,1}+1 \right)
		+x_{3,n,2}\left( x_{3,n,1}+1 \right)
		+x_{2,n+\hat 1,2}\left( x_{2,n+\hat 1,1}+x_{2,n+\hat 1,2} \right)
		+x_{3,n+\hat 1,2}\left( x_{3,n+\hat 1,1}+x_{3,n+\hat 1,2} \right)
	}
\end{align}
\end{widetext}
where the coordinates
in $G$
are unified as in the previous subsection.
Note that
from Eq.~(\ref{eq:modT}) and ~(\ref{eq:xf})
a new constraint for the updated indices is given by
\begin{widetext} 
\begin{equation}
	p_{ l^{\rm{f}}_{n^\star} }\equiv
	\left[
		\sum_{s=1}^2
		\left( x_{1,n^\star,s} +x_{1,n^\star-\hat 1^\star,s} \right)
		+\sum_{i=2}^3
		\left( x^{\rm{f}}_{i,n^\star}+x^{\rm{f}}_{i,n^\star-\hat i} \right)
	\right]
	\bmod 2=0.
	\label{eq:modTstar}
\end{equation}
\end{widetext}

By combining the renormalization procedures for the bosonic and the Grassmann parts,
the coarse-grained tensor is expressed as
\begin{equation}
	\mathcal T^\star_{ l_{n^\star} }=T^\star_{ l_{n^\star} } G^\star_{ l^{\rm{f}}_{n^\star} },
\end{equation}
where the indices $l_{n^\star}$ are defined as 
\begin{equation}
	l_{n^\star}= \left( l^{\rm{b}}_{n^\star},l^{\rm{f}}_{n^\star} \right)
\end{equation}
and the coarse-grained bosonic tensor which has the new indices arise from the Grassmann part is given by
\begin{align}
	T^\star_{ l_{n^\star} }
	=T^{\star,\rm{b}}_{ l^{\rm{b}}_{n^\star} } \delta_{ p_{ l^{\rm{f}}_{n^\star} },0 }
	\label{eq:Tstar}
\end{align}
with the Kronecker delta to satisfy Eq.~(\ref{eq:modTstar}).

\section{Calculational method for fermionic Green functions}
\label{sec:Gfunction}
\subsection{Chiral condensate}

We first explain a method to calculate the chiral condensate defined as
\begin{gather}
	\left< \bar\psi\psi \right>
	=\frac{Z_I}{Z}, \\
	Z_I
	=\frac{1}{V}\int \mathcal D\psi \mathcal D\bar\psi e^{-S}
	\left( \sum_{n,i} \bar\psi_{n,i}\psi_{n,i} \right),
	\label{eq:chiralCondensate}
\end{gather}
where $V$ is the system volume.
Thanks to the translational invariance,
Eq.~(\ref{eq:chiralCondensate}) is rewritten as
\begin{equation}
	Z_I
	=\int \mathcal D\psi \mathcal D\bar\psi e^{-S}
	\left( \sum_i \bar\psi_{n_0,i}\psi_{n_0,i} \right).
\end{equation}
Due to the presence of $\bar\psi_{n_0}\psi_{n_0}$, the bosonic tensor on the site $n_0$ is modified.
Now we introduce ``impure tensor"
\footnote{
	The coarse-graining for the impure tensor in the framework of TRG is discussed in
	Ref~\cite{impureTRG,nakamoto}.} 
and define the bosonic part in it as $I$.
Then the impure tensor with the Grassmann part is defined as
\begin{equation}
	\mathcal I_{ l_{n_0} }= I_{ l_{n_0} } G_{ l_{n_0} }.
\end{equation}
Then Eq.~(\ref{eq:chiralCondensate}) is expressed as
\begin{equation}
	Z_I	=\sum_{\{ x \}}	\mathcal I_{ l_{n_0} } \prod_{n\neq n_0}\mathcal T_{ l_n }.
	\label{eq:impureTensor}
\end{equation}
In the following the renormalization algorithm is formulated based on this expression.

\begin{figure}
	\centering
	\includegraphics[scale=0.4]{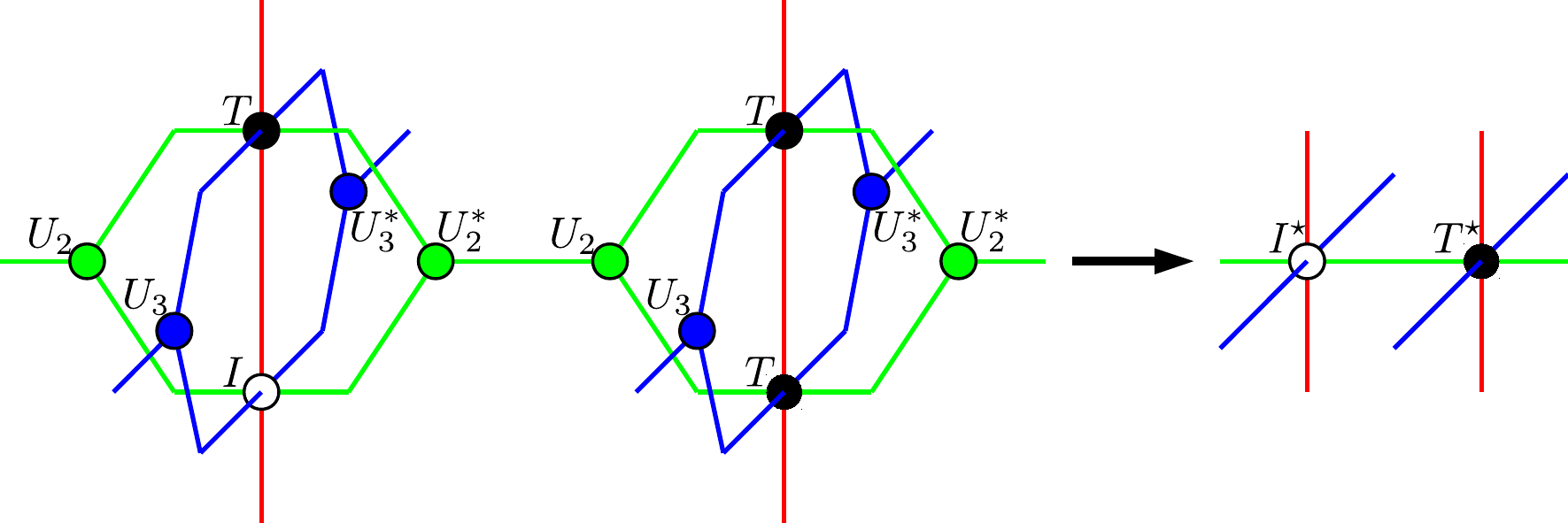}
	\caption{Coarse-graining procedure for the impure tensor.}
	\label{fig:impureCC}
\end{figure}

First of all, the renormalization procedure for the normal tensors is explained in Sec.~\ref{sec:GHOTRG}.
On the other hand, in case of the the impure tensor,
the contracted tensor of Eq.~(\ref{eq:tensorM}) is modified as
\begin{equation}
	M^{\rm{impure}}_{ m_{n_0} }
	=\sum_{ x_{n_0} } T_{ l_{n_0+\hat 1} } I_{ l_{n_0} } \sigma_{ l_{n_0+\hat 1} l_{n_0} }.
\end{equation}
With the use of the isometries that are the same as in the normal tensors,
we obtain the following new bosonic tensor:
\begin{multline}
	I^{\star,\rm{b}}_{ l^{\rm{b}}_{n_0^\star} }
	=\sum_{ x_{2,n_0}^\pm,x_{3,n_0}^\pm } M^{\rm{impure}}_{ m_{n_0} } \\
	\times \prod_{i=2}^3
	\left( U_i \right)^\ast_{ x_{i,n_0}^+ x^{\rm{b}}_{i,n_0^\star} }
	\left( U_i \right)_{ x_{i,n_0}^- x^{\rm{b}}_{i,n_0^\star-\hat i} }.
\end{multline}
Figure~\ref{fig:impureCC} gives a schematic view of the coarse-graining procedure for the impure tensor.
The renormalization procedure for the Grassmann part is similar to that for the normal tensors.
Thus the coarse-grained impure tensor is expressed as
\begin{gather}
	\mathcal I^\star_{ l_{n_0^\star} }
	=I^\star_{ l_{n_0^\star} } G^\star_{ l^{\rm{f}}_{n_0^\star} }, \\
	I^\star_{ l_{n_0^\star} }
	=I^{\star,\mathrm{b}}_{ l^{\rm{b}}_{n_0^\star} } \delta_{ p^{\rm{f}}_{n_0^\star },0 }.
\end{gather}

\subsection{Fermion two-point correlation functions}
We now explain a calculational method for the fermion two-point correlation function defined as
\begin{align}
	C_{s_1s_2}(n_1,n_2)= \frac{Z_{s_1s_2}(n_1,n_2)}{Z}
\end{align}
with
\begin{align}	
	Z_{s_1s_2}(n_1,n_2)= \int \mathcal D\psi \mathcal D\bar\psi e^{-S} \bar\psi_{n_1,s_1}\psi_{n_2,s_2}.
	\label{eq:Z2}
\end{align}
In this case, the tensors on the site $n_1$ and $n_2$ become the impure tensors,
which are denoted by
\begin{gather}
	\bar{\mathcal I}_{ l_{n_1} }= \bar I_{ l_{n_1} } G_{ l_{n_1} }, \\ 
	\mathcal I_{ l_{n_2} }= I_{ l_{n_2} } G_{ l_{n_2} },
\end{gather}
respectively.
Inclusion of odd numbers of fermions at the site $n_1$ and $n_2$  
modifies the constraint described in Eq.~(\ref{eq:modT}) as
\begin{align}
	\left[ \sum_{i=1}^3 \sum_{s=1}^2 \left( x_{i,n,s}+x_{i,n-\hat i,s} \right) \right]
	\bmod 2=1.
	\label{eq:modI}
\end{align}
Then Eq.~(\ref{eq:Z2}) is expressed as
\begin{align}
	Z_{s_1s_2} =\sum_{\{x\}} \bar{\mathcal I}_{ l_{n_1} } \mathcal I_{ l_{n_2} }
	\prod_{n\neq n_1,n_2} \mathcal T_{ l_n }.
\end{align}

In calculation of the fermion two-point correlation functions,
the renormalization procedures are classified into three types according to the positions of the impure tensors.
We explain them in order.

\subsubsection{$n_1+\hat 1=n_2$}
\begin{figure}
	\centering
	\includegraphics[scale=0.4]{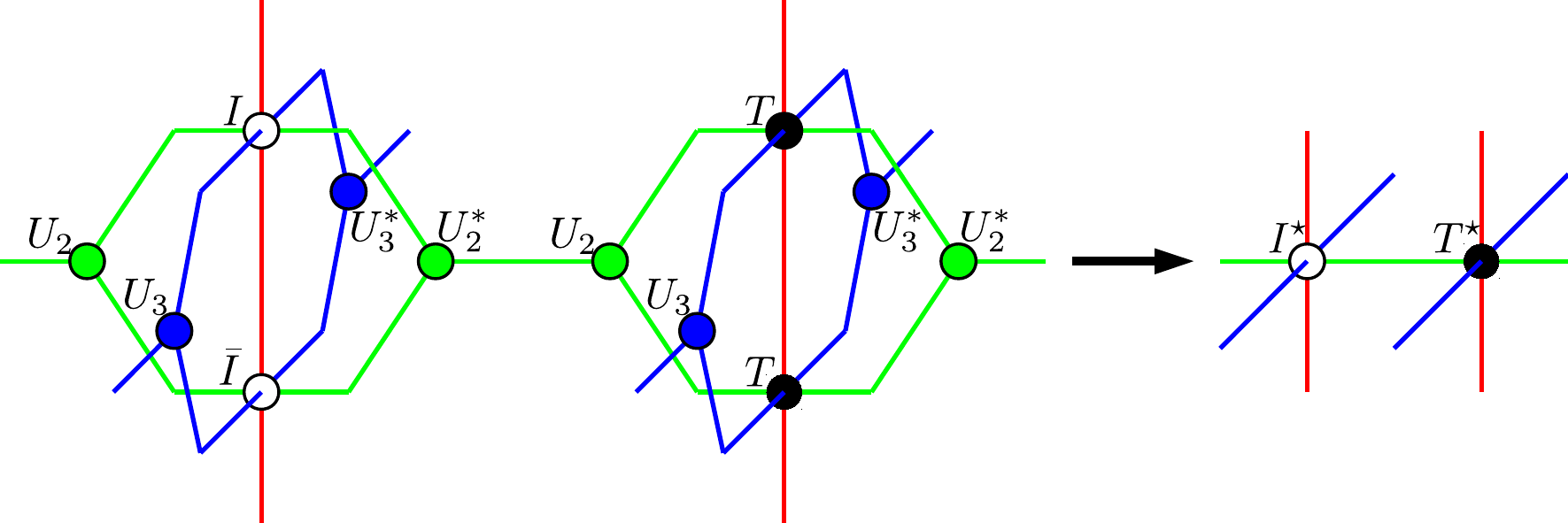}
	\caption{Coarse-graining procedure for two impure tensors in the case of $n_1+\hat 1=n_2$.}
	\label{fig:impureGreen1}
\end{figure}
In this case two impure tensors are merged into one impure tensor as shown in Fig.~\ref{fig:impureGreen1}.
The contracted tensor is obtained by
\begin{equation}
	M^{\mathrm{impure}}_{ m_{n_1} }
	=\sum_{ x_{1,n_1} } I_{ l_{n_2} } \bar I_{ l_{n_1} } \sigma^\prime_{ l_{n_2} l_{n_1} }.
\end{equation}
where $\sigma^\prime$ is the sign factor modified by Eq.~(\ref{eq:modI}) as follows:
\begin{equation}
	\sigma^\prime_{ l_{n_2} l_{n_1} }
	=(-1)^{ \sum_{i=2}^3 \sum_{s=1}^2 x_{i,n_2-\hat i,s} } \sigma_{ l_{n_2} l_{n_1} }.
\end{equation}
We use the isometries which are the same as that for the normal tensors.
After the two impure tensors are merged,
we obtain a tensor network with a single impure tensor described in the previous subsection.

\subsubsection{$n_1\pm\hat 2=n_2$ or $n_1\pm\hat 3=n_2$}
\begin{figure}
	\centering
	\includegraphics[scale=0.4]{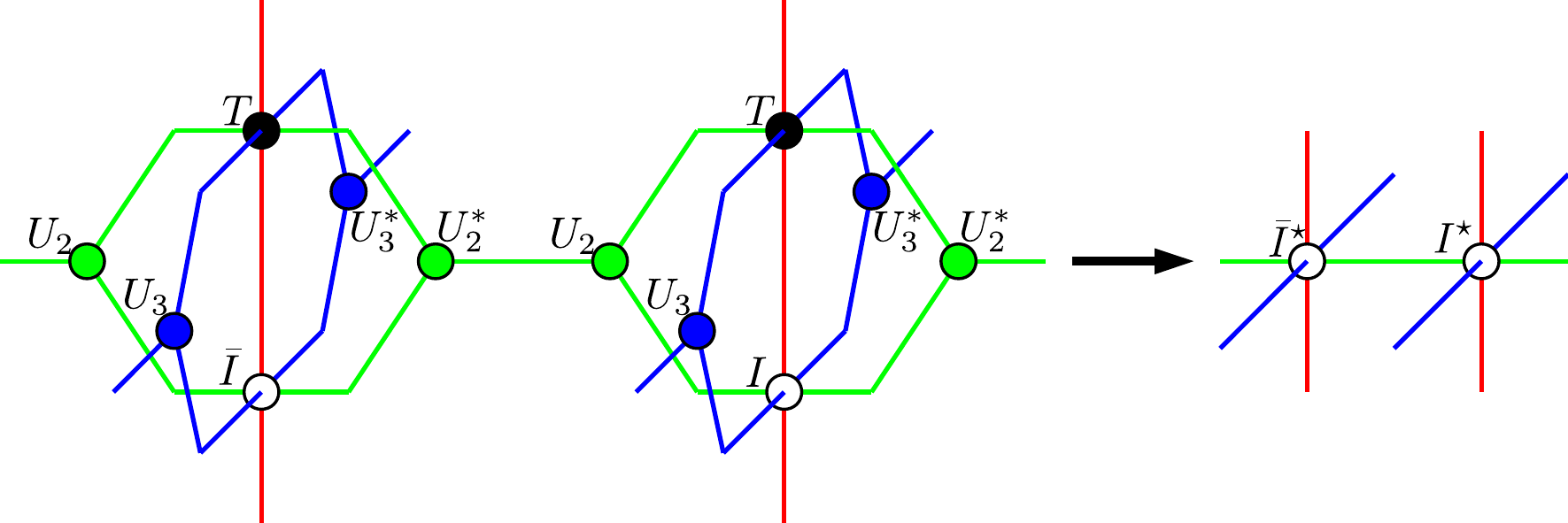}
	\caption{Coarse-graining procedure for two impure tensors in the case of $n_1\pm\hat 2=n_2$.}
	\label{fig:impureGreen2}
\end{figure}

We exemplify the case of $n_1+\hat 2=n_2$ as depicted in Fig.~\ref{fig:impureGreen2}.
For each impure tensor we construct the corresponding contracted tensor: 
\begin{gather}
	\bar M^{\mathrm{impure}}_{ m_{n_1} }
	=\sum_{x_{1,n_1}} T_{ l_{n_1+\hat 1} } \bar I_{ l_{n_1} } \sigma^\prime_{ l_{n_1+\hat 1} l_{n_1} }, \\
	M^{\mathrm{impure}}_{ m_{n_2} }
	=\sum_{x_{1,n_2}} T_{ l_{n_2+\hat 1} } I_{ l_{n_2} } \sigma^\prime_{ l_{n_2+\hat 1} l_{n_2} }.
\end{gather}
The unitary matrices $U_2,U_3$ are obtained by the renormalization procedure for the normal tensors.
In addition, we apply the eigenvalue decomposition to $M^+$ and $M^-$ defined by
\begin{align}
	M^{\mathrm{impure},+}_{x_{2,n_1}^+,x_{2,n_1}^{+\prime}}
	=\sum_{ a,c,d,e,f }
	\bar M^{\mathrm{impure}}_{ a x_{2,n_1}^+ cdef }
	\bar M^{\mathrm{impure}}_{ a x_{2,n_1}^{+\prime} cdef },
	\\
	M^{\mathrm{impure},-}_{x_{2,n_2}^-,x_{2,n_2}^{-\prime}}
	=\sum_{ a,b,c,d,f }
	M^{\mathrm{impure}}_{ abcd x_{2,n_2}^- f }
	M^{\mathrm{impure}}_{ abcd x_{2,n_2}^{-\prime} f }
\end{align}
and obtain a unitary matrix $U^\pm$ and the eigenvalues $\lambda^\pm$ and $\epsilon^\pm$.
in the same manner of Eq.~(\ref{eq:evd}) and Eq.~(\ref{eq:epsilon}).
If $\epsilon^+ < \epsilon^-$ then $W=U^+$ is adopted, and vice versa.
The new bosonic tensors $\bar I^\star,I^\star$ are obtained by using $U_2,U_3$ and $W$:
\begin{widetext}
\begin{gather}
	\bar I^{\star,\rm{b}}_{ l^{\rm{b}}_{n_1^\star } }
	=\sum_{ x^\pm_{2,n_1},x^\pm_{3,n_1} }
	\bar M^{\mathrm{impure}}_{ m_{n_1} }
	W^\ast_{ x^+_{2,n_1} x^{\rm{b}}_{2,n_1^\star} }
	(U_3)^\ast_{ x^+_{3,n_1} x^{\rm{b}}_{3,n_1^\star} }
	\prod_{i=2}^3 (U_i)_{ x^-_{i,n_1} x^{\rm{b}}_{i,n_1^\star-\hat i} },
	\\
	I^{\star,\mathrm{b}}_{ l^{\rm{b}}_{n_2^\star } }
	=\sum_{ x^\pm_{2,n_1},x^\pm_{3,n_1} }
	M^{\mathrm{impure}}_{ m_{n_2} }
	\prod_{i=2}^3 (U_i)_{ x^+_{i,n_1} x^{\rm{b}}_{i,n_1^\star} }
	W_{ x^-_{2,n_2} x^{\rm{b}}_{2,n_2^\star-\hat 2} }
	(U_3)_{ x^-_{3,n_2} x^{\rm{b}}_{3,n_2^\star-\hat 3} }.
\end{gather}
\end{widetext}
The rest of the procedure is similar to that for the normal tensors. 

\subsubsection{Other cases}
This is the case that
two impure tensors are renormalized independently. 
The contracted tensors for impure tensors are given by
\begin{gather}
\bar M^{\mathrm{impure}}_{ m_{n_1} }
=\sum_{x_{1,n_1}} T_{ l_{n_1+\hat 1} } \bar I_{ l_{n_1} } \sigma^\prime_{ l_{n_1+\hat 1} l_{n_1} }, \\
M^{\mathrm{impure}}_{ m_{n_2} }
=\sum_{x_{1,n_2}} T_{ l_{n_2+\hat 1} } I_{ l_{n_2} } \sigma^\prime_{ l_{n_2+\hat 1} l_{n_2} }.
\end{gather}
The isometries are obtained by the renomalization procedure for the normal tensors.
The rest of the procedure is similar to that for the normal tensors. 

\section{Numerical results}
\label{sec:results}
In this section, we present the numerical results.
The anti-periodic boundary condition is imposed for one of the directions.
We employ it to test the calculational method of the fermionic Green functions on a $256^{3}$ lattice.

\subsection{Free energy}
\begin{figure}
	\centering
	\includegraphics[scale=0.5]{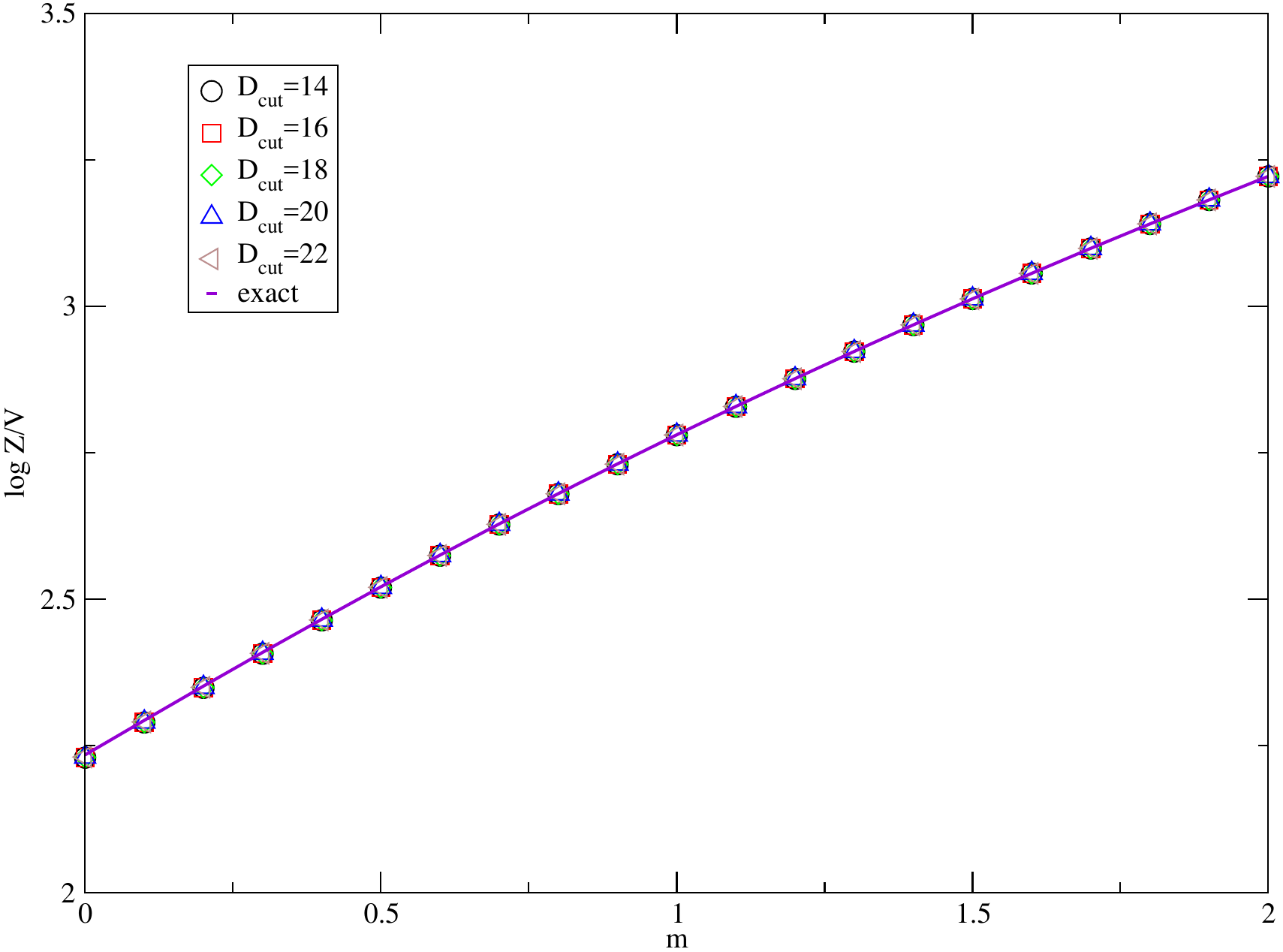}
	\caption{Free energy as a function of $m$.}
	\label{fig:freeEnergy}
\end{figure}

\begin{figure}
	\centering
	\includegraphics[scale=0.5]{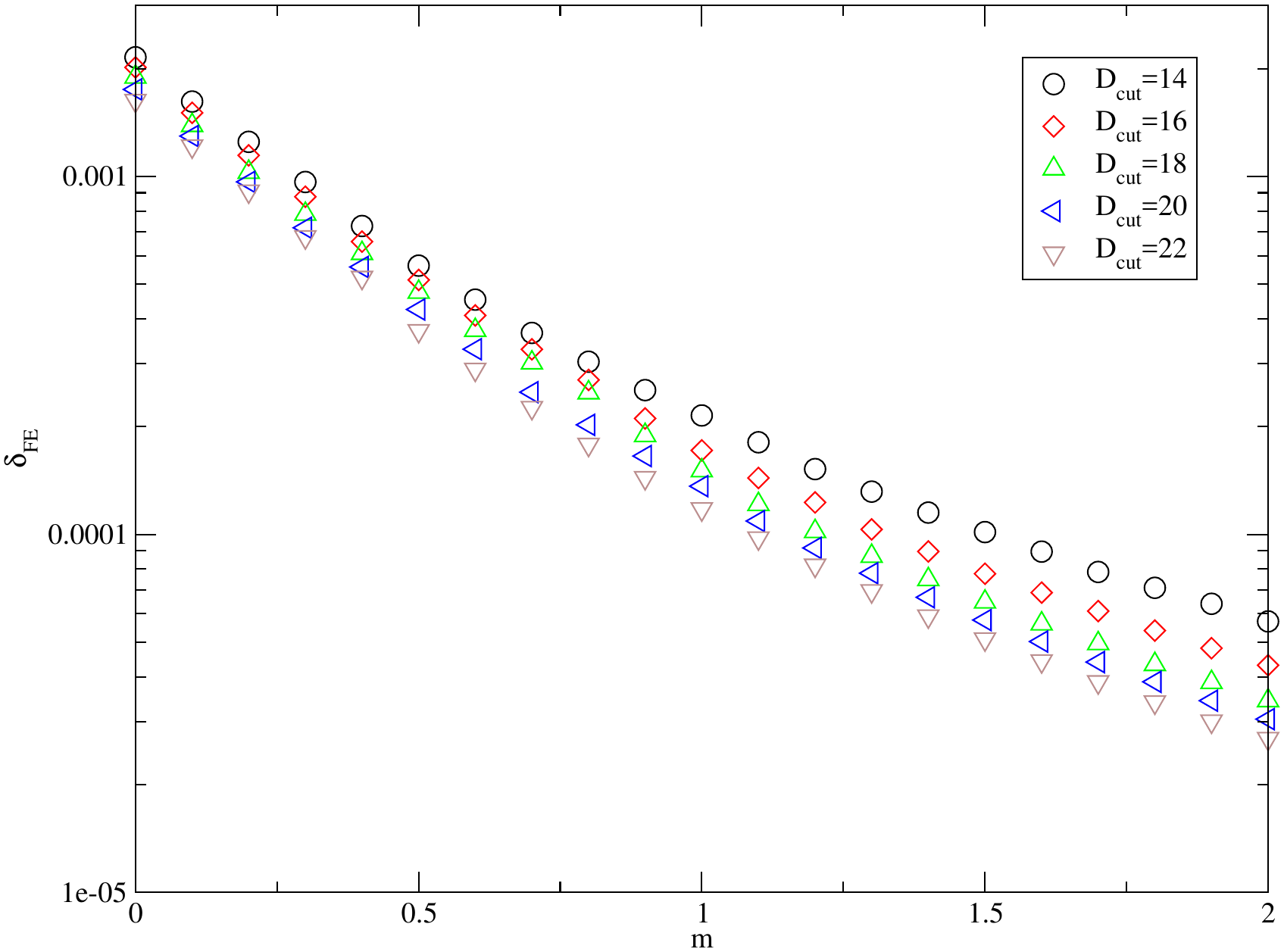}
	\caption{Error of free energy defined by Eq.~(\ref{eq:delta_fe}) as a function of $m$.}
	\label{fig:deltaFreeEnergy}
\end{figure}

We first present the results for the free energy of three-dimensional free Wilson fermion system.
Figure~\ref{fig:freeEnergy} compares the numerical results with the exact values for the free energy as a function of the fermion mass $m$. The latter is obtained by an analytical calculation in the momentum spaces.
It is hard to detect the deviation between the measured values  with $D_{\rm cut}\ge 14$ and the exact ones. In Fig.~\ref{fig:deltaFreeEnergy} we directly plot the magnitude of the error defined by
\begin{align}
	\delta_{\mathrm{FE}}= \frac{|\ln Z_{\mathrm{exact}}-\ln Z(D_{\mathrm{cut}})|}{|\ln Z_{\mathrm{exact}}|},
\label{eq:delta_fe}
\end{align}
which clarify the difference between the numerical results with the exact values.
We observe that the error decreases monotonically in the range of $0\le m\le 2$ as $D_{\mathrm{cut}}$ increases from 14 to 22. It is also noted that the error diminishes monotonically as the fermion mass becomes larger.
The maximal error is about 0.2\% at $m=0$ and the minimal one is about 0.02\% at $m=2$.


\subsection{Chiral condensate}
\begin{figure}
	\centering
	\includegraphics[scale=0.5]{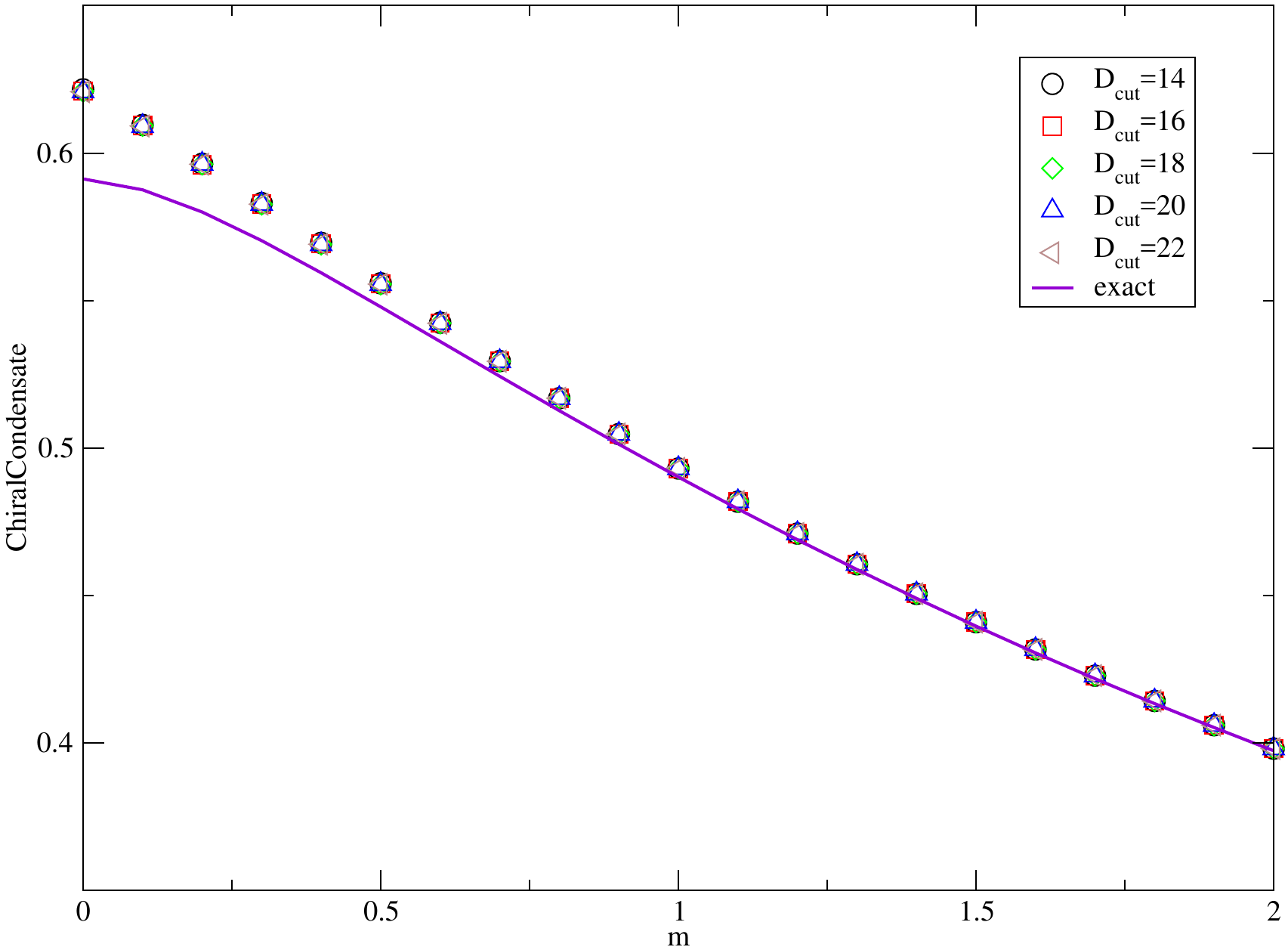}
	\caption{Chiral condensate as a function of $m$.}
	\label{fig:chiralCondensate}
\end{figure}

\begin{figure}
	\centering
	\includegraphics[scale=0.5]{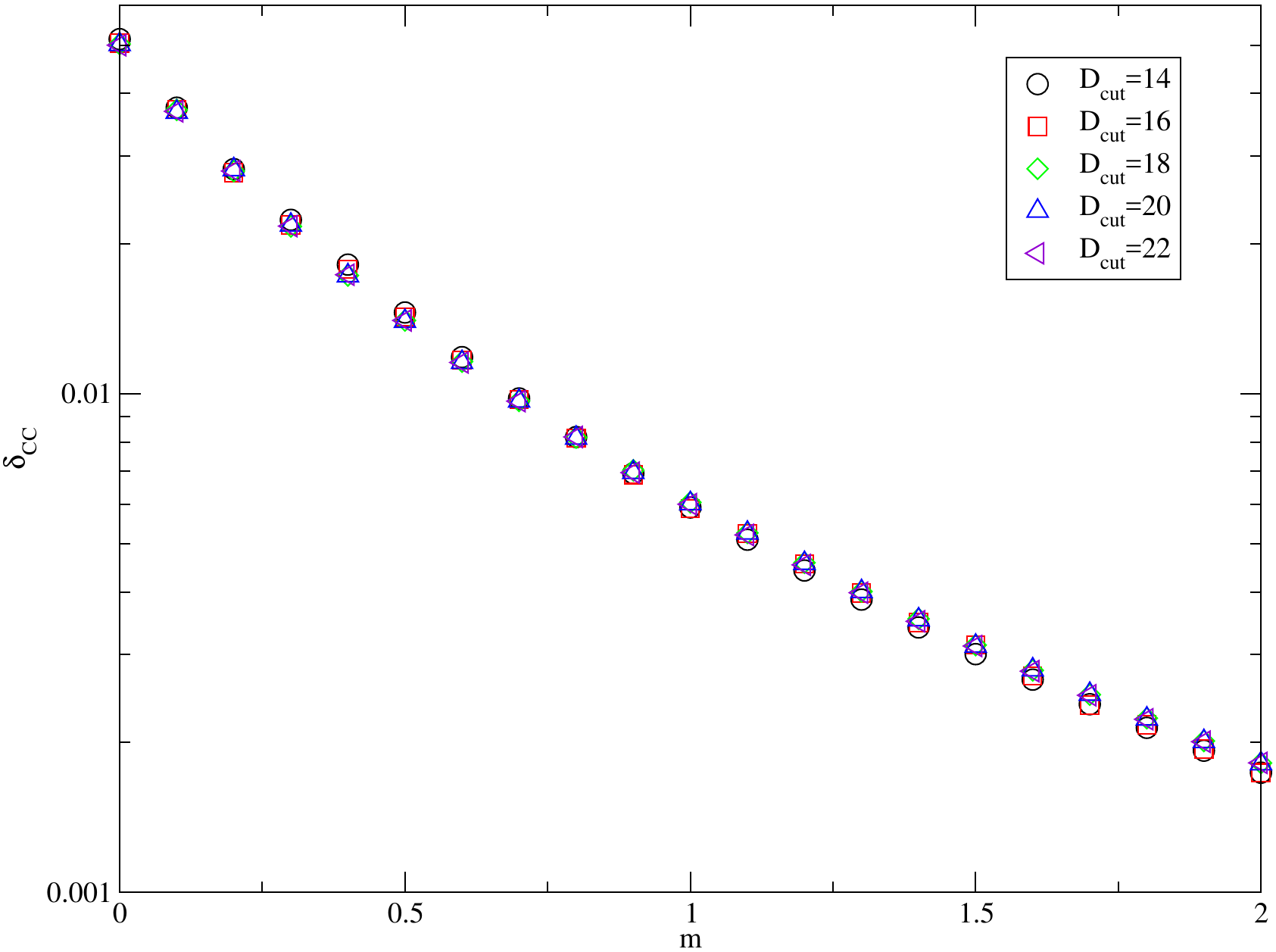}
	\caption{Error of chiral condensate defined by Eq.~(\ref{eq:delta_cc}) as a function of $m$.}
	\label{fig:errorChiralCondensate}
\end{figure}

We summarize the results for the chiral condensate in Figs.~\ref{fig:chiralCondensate} and \ref{fig:errorChiralCondensate}.
The former plots the numerical results as a function of the fermion mass $m$ together with the exact values. We find a clear deviation between them near $m=0$. The latter shows the magnitude of the error defined by
\begin{align}
	\delta_{\mathrm{CC}}
	=\frac{|\left<\bar\psi\psi\right>_{\mathrm{exact}}-\left<\bar\psi\psi\right>(D_{\mathrm{cut}})|}
	{|\left<\bar\psi\psi\right>_{\mathrm{exact}}|}.
\label{eq:delta_cc}
\end{align}
The error reduces monotonically as the fermion mass increases. This is a similar tendency as observed for the error of the free energy.
The maximal error is about 5\% at $m=0$ and the minimal one is about 0.1\% at $m=2$.
It should be noted that the error does not change so much against the increment of $D_{\mathrm{cut}}$.
The reason may be that we use the common unitary matrices, which are optimized for the normal tensors but not for the impure tensors, in the coarse-graining of both tensors.


\subsection{Fermion two-point correlation functions}

\begin{figure}
	\centering
	\includegraphics[scale=0.5]{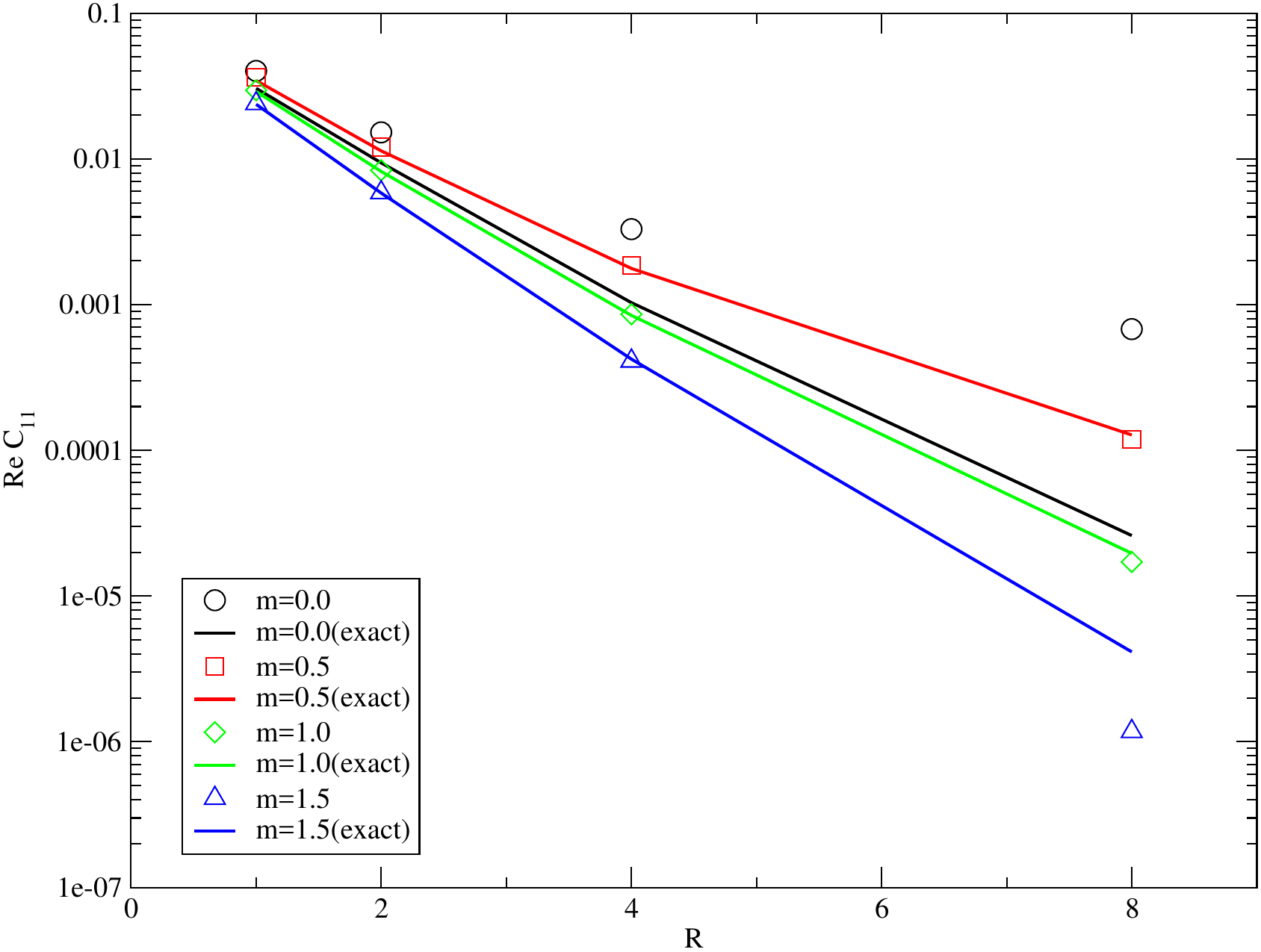}
	\caption{Real part of $C_{11}\left(n_1,n_2\right)$.}
	\label{fig:bp1p1}
\end{figure}
\begin{figure}
	\centering
	\includegraphics[scale=0.5]{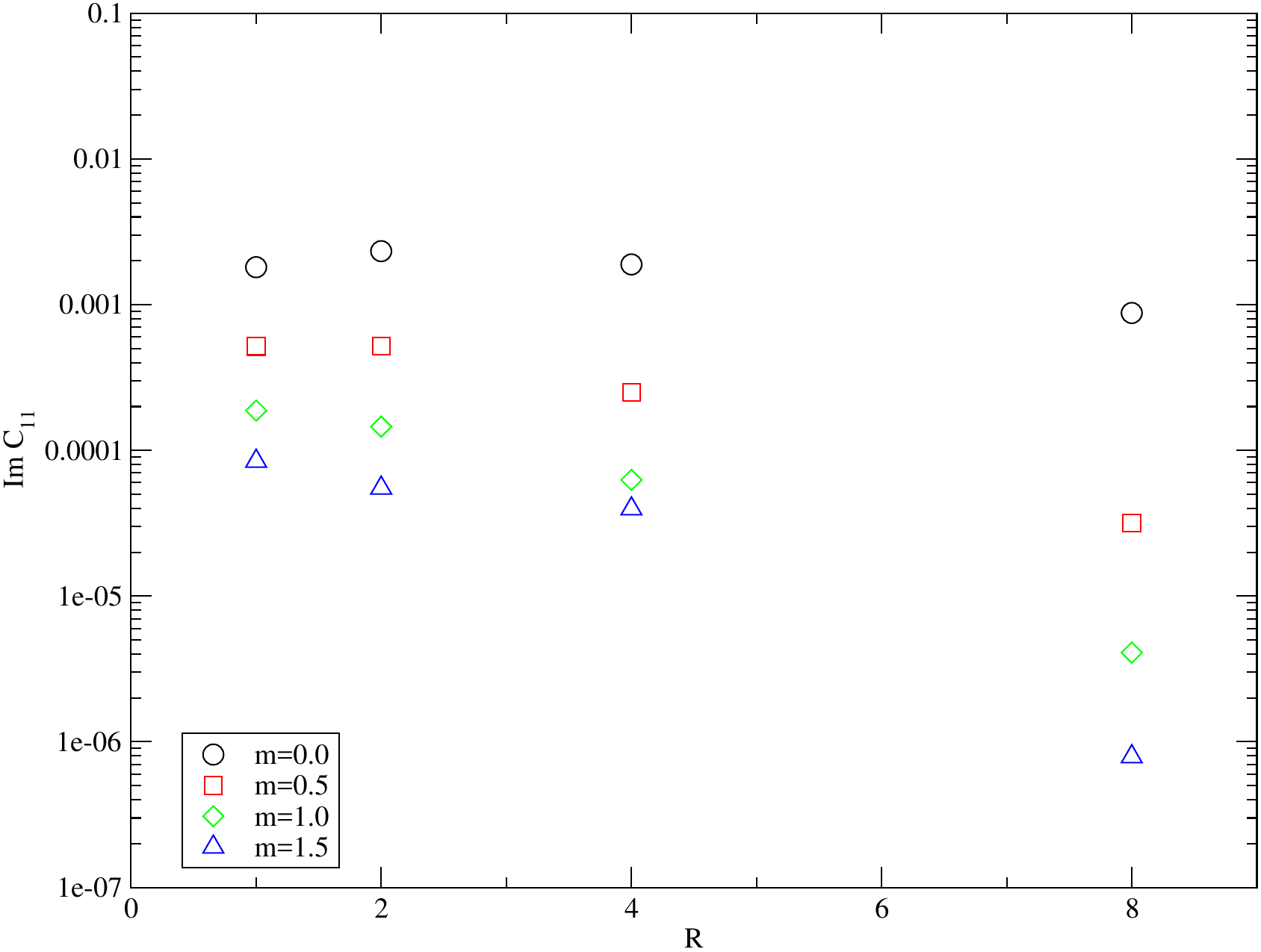}
	\caption{Imaginary part of $C_{11}\left(n_1,n_2\right)$.}
	\label{fig:ibp1p1}
\end{figure}
\begin{figure}
	\centering
	\includegraphics[scale=0.5]{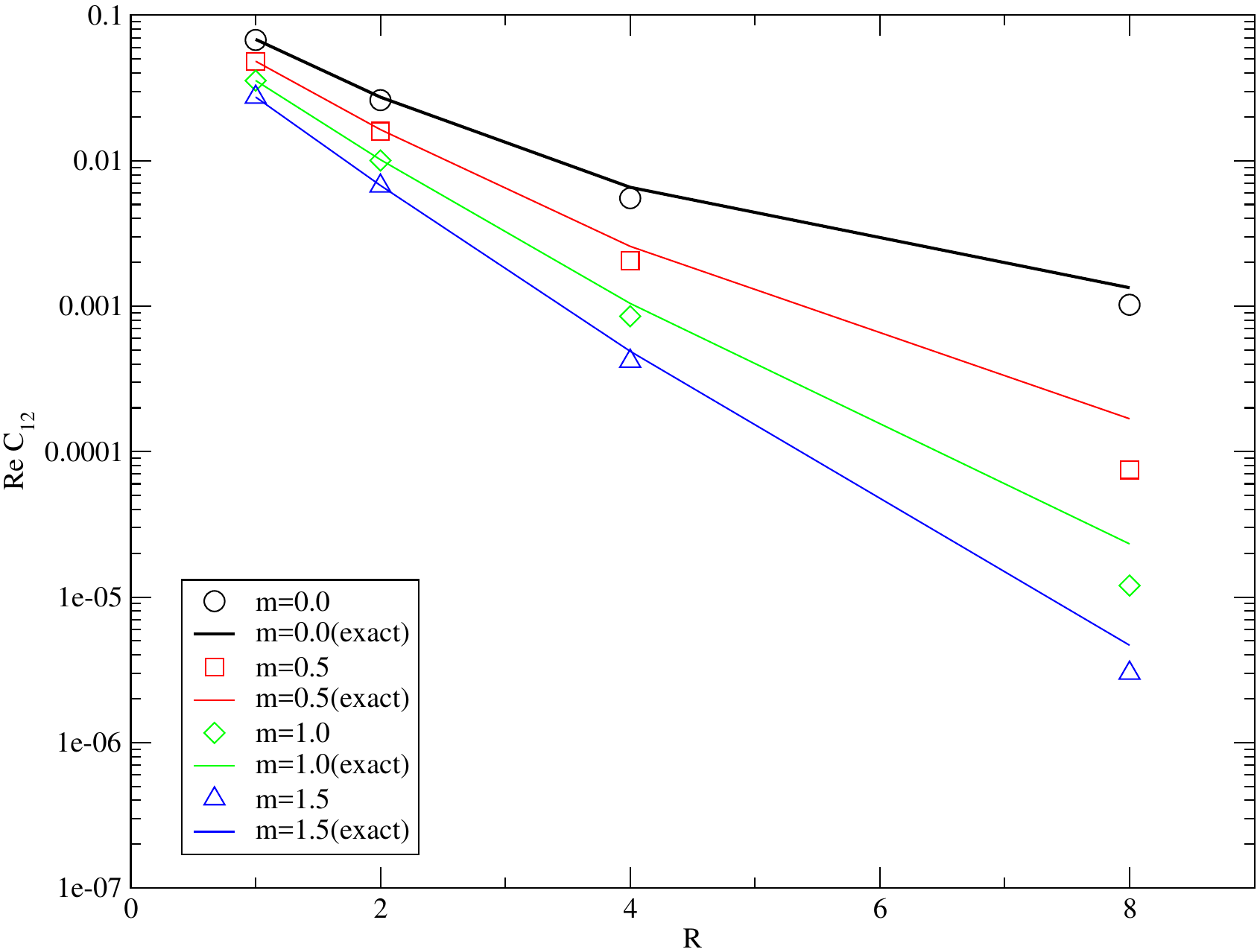}
	\caption{Real part of $C_{12}\left(n_1,n_2\right)$.}
	\label{fig:bp1p2}
\end{figure}
\begin{figure}
	\centering
	\includegraphics[scale=0.5]{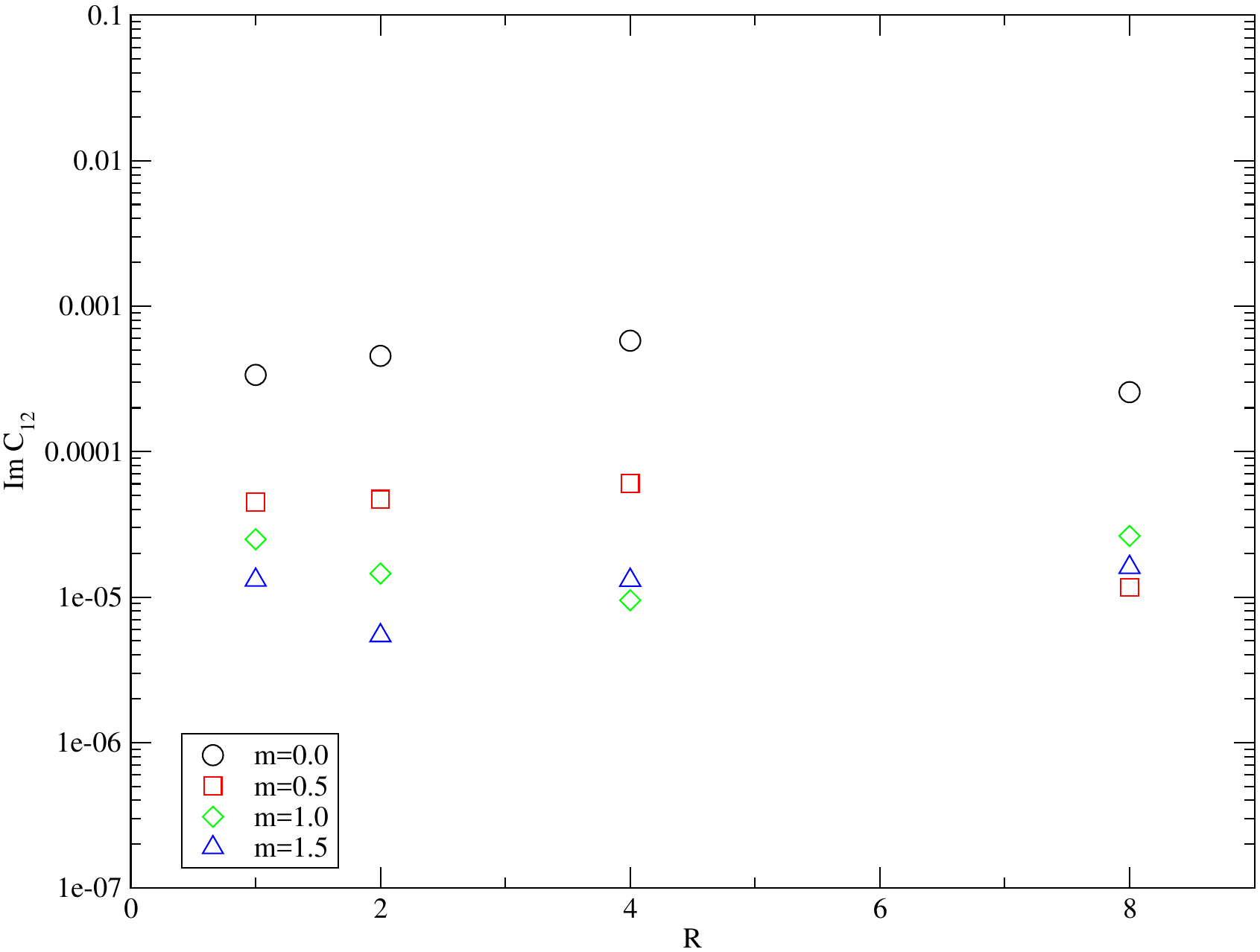}
	\caption{Imaginary part of $C_{12}\left(n_1,n_2\right)$.}
	\label{fig:ibp1p2}
\end{figure}

We consider $C_{11}\left(n_1,n_2\right)$ and $C_{12}\left(n_1,n_2\right)$ as representative cases.
Figures~\ref{fig:bp1p1}, \ref{fig:ibp1p1}, \ref{fig:bp1p2} and \ref{fig:ibp1p2} show real and imaginary parts of
$C_{11}\left(n_1,n_2\right)$ and $C_{12}\left(n_1,n_2\right)$, respectively, as a function of $R$, which is defined by $n_2=n_1+R\cdot\hat 1$.
In Figs.~\ref{fig:bp1p1} and \ref{fig:bp1p2}
we find that the accuracy of real part becomes worse on $m=0$ 
as observed for the free energy and the chiral condensate.
Another important finding is that the errors become relatively large at larger $R$ for all the choices of $m$.
Here we point out that the magnitude of the imaginary part may give us an estimate of the magnitude of the error for the real part, since the imaginary part is expected to vanish theoretically.
Figures~\ref{fig:ibp1p1} and \ref{fig:ibp1p2}, in fact, show that the magnitude of the real part at $R=8$ are comparable with that of the imaginary part.
This indicates that the signal-to-noise ratio of the real part at the large distance (around $R=8$) may be of order of one.
In our case, where the correlation length is small, it is difficult to accurately compute the correlation function at large separation.
For the case of the larger correlation length, however,
one may measure the long separated correlation function, whose signal-to-noise ratio is expected to be better. 

\section{Summary and outlook}
\label{sec:summary}
We have developed a method to calculate fermionic Green functions in the framework of GHOTRG using the impure tensors. The validity of the method is tested by calculating the chiral condensate and the fermion two-point correlation functions in the three-dimensional free Wilson fermion system on a $256^3$ lattice, where we employ $0\le m\le 2$ for the fermion mass parameter and  $14\le D_{\rm cut}\le 22$ for the tensor indices. In most cases we find good consistency between the numerical results and the exact values: their deviation is at most a few \% level. Only exception is the fermion two-point correlation functions at a long distance. It may be due to the difficulty of evaluating tiny correlations left at a long distance. 

The method developed in this paper can be applied to any physical systems which contain fermions.
We plan to apply it to more complicated systems including gauge interactions.
It is also important to improve the accuracy of the method. It could be effective to use an idea of the higher-order second renormalization group (HOSRG)~\cite{hotrg,srg} for the formulation of impure tensors.

\begin{acknowledgments}
This research used computational resources of HA-PACS and COMA provided by Interdisciplinary Computational Science Program in Center for Computational Sciences, University of Tsukuba.
This research is supported by the Ministry of Education, Culture, Sports, Science and Technology (MEXT) as ``Exploratory Challenge on Post-K computer (Frontiers of Basic Science: Challenging the Limits)'' and also in part by Grants-in-Aid for Scientific Research from MEXT (Nos. 15H03651).
\end{acknowledgments}

\appendix
\def\thesection{\Alph{section}}
\section{Block diagonalization}
We explain the block diagonalization for the bosonic tensors.
Let us consider the following matrix representation for a given bosonic tensor
\begin{align}
	T^{\mathrm{matrix}}_{r,c}=T_{ x_{1,n}x_{2,n}x_{3,n}x_{1,n-\hat 1}x_{2,n-\hat 2}x_{3,n-\hat 3} }
\end{align}
with
\begin{align}
	r&=x_{1,n}\otimes x_{2,n}\otimes x_{3,n}, \\
	c&=x_{1,n-\hat 1}\otimes x_{2,n-\hat 2}\otimes x_{3,n-\hat 3}.
\end{align}
Thanks to the condition of Eq.~(\ref{eq:modT}) 
the matrix can be block-diagonalized as
\begin{align}
	T^{\mathrm{matrix}}=
	\begin{array}{rccll}
		                  &p_{\mathrm{col}}=0 &p_{\mathrm{col}}=1 &                   &          \\
		\ldelim({2}{3mm}[]&T^{\mathrm{even}}  &0                  &\rdelim){2}{3mm}[] &p_{\mathrm{row}}=0\\
		                  &0                  &T^{\mathrm{odd}}   &                   &p_{\mathrm{row}}=1\\
	\end{array}
\end{align}
with
\begin{align}
	p_{\mathrm{row}}
	&=\left( \sum_{i=1}^3 \sum_{s=1}^2 x_{i,n,s} \right) \bmod 2,
	\\
	p_{\mathrm{col}}
	&=\left( \sum_{i=1}^3 \sum_{s=1}^2 x_{i,n-\hat i,s} \right) \bmod 2.
\end{align}

Similarly,
the contracted tensor of Eq.~(\ref{eq:tensorM}) can be also block-diagonalized. 
Thus the eigenvalue decomposition of Eq.~(\ref{eq:evd}) is independently applied to the even and odd blocks yielding the associated unitary matrices $U_i^{\mathrm{even}}$ and $U_i^{\mathrm{odd}}$.
The block-diagonalized unitary matrix is expressed as
\begin{align}
	U_i^{\mathrm{matrix}}=
	\begin{array}{rccll}
		&x_{i,n^\star,\mathrm{f}}=0 &x_{i,n^\star,\mathrm{f}}=1 & & \\
		\ldelim({2}{3mm}[] &U_i^{\mathrm{even}} &0 &\rdelim){2}{3mm}[] & p_n=0\\
		&\underbrace{0} &\underbrace{U_i^{\mathrm{odd}}} & &p_n=1,\\
		&D_{\mathrm{even}} &D_{\mathrm{odd}} & &
	\end{array}
\end{align}
with
\begin{align}
	p_n=\left( \sum_{s=1}^2 x_{i,n,s}+x_{i,n-\hat i,s} \right) \bmod 2
\end{align}
where $D_{\mathrm{even}}+D_{\mathrm{odd}}=D_{\mathrm{cut}}$.

This property can be used to reduce the computational and memory costs by dealing with only non-trivial blocks.
Moreover, the blocks satisfy the condition of Eq.~(\ref{eq:modTstar}) so that
it is not necessary to extend the fermionic indices for the bosonic tensor as Eq.~(\ref{eq:Tstar}).

\end{document}